\documentclass{PoS}

\title{Dual Superconductivity in $G_2$ group}

\ShortTitle{Dual Superconductivity in $G_2$ group}

\author{\speaker{Guido Cossu}\\
        Scuola Normale Superiore \&  INFN, Pisa\\
        E-mail: \email{g.cossu@sns.it}}
\author{Massimo D'Elia\\
        Dipartimento di Fisica \& INFN, Genova\\
        E-mail: \email{delia@ge.infn.it}}
\author{Adriano Di Giacomo and Claudio Pica\\
        Dipartimento di Fisica \& INFN, Pisa\\
        E-mail: \email{digiacomo@df.unipi.it},~  \email{pica@df.unipi.it}}
\author{Biagio Lucini\\
        Department of Physics, University  of Wales, Swansea\\
        E-mail: \email{b.lucini@swansea.ac.uk}}

\abstract{We investigate the dual superconductivity mechanism in the exceptional
group $G_2$. This is a centerless group (no 't Hooft flux vortices are allowed)
 and we check for the presence of a magnetic monopole condensate in the confined phase by measuring on the lattice a disorder parameter related to the vacuum
expectation value of an operator carrying magnetic charge. The behaviour of the disorder parameter is consistent with the dual superconductor picture. A first step of an analysis on the thermodynamical properties of the theory is conducted by mean of this operator.}

\FullConference{XXIVth International Symposium on Lattice Field Theory\\
                July 23-28, 2006\\
                Tucson, Arizona, USA}

\begin{document}

\section{Introduction and Motivation}
Understanding confinement is one of the main issues in non perturbative QCD. 
In this respect identifying the degrees of freedom responsible of quark confinement is essential. Many choices are investigated in literature, between them center vortices and monopoles (both proposed by 't Hooft) are the most popular. The deconfinement transition in pure SU(3) gauge theory is triggered by the spontaneous breaking of center symmetry, this could suggest the direct relation between confinement and presence of a non trivial center. It's therefore an interesting question whether centerless groups have a confinement-deconfinement phae transition. In fact the $SO(3) = SU(2)/Z_2$ has been extensively studied on the lattice (see for example \cite{Datta:1999np, Barresi:2004qa, deForcrand:2002vs}). Moreover the exceptional group $G_2$ is without 't Hooft center vortices and is the simplest one, that's why we focused on it.

Another point of view concerning the confinement mechanism is the so called \emph{dual superconductor picture}, the subject of this work. As in an ordinary superconductor magnetic fields are squeezed in flux tubes (Abrikosov filaments) connecting magnetic charges due to condensation of Cooper pairs, quarks could be kept together by strings of (chromo)electric flux tubes due to condensation of magnetic objects. According to this picture, the QCD vacuum,  is a condensate of magnetically charged fields confining (chromo)electrically charged particles.  

The presence of strings connecting static quarks has been proved in several papers. A way of probing the vacuum state looking for the Abelian magnetic fields has been proposed and studied by the Pisa group in a series of papers \cite{DiGiacomo:1999fa, DiGiacomo:1999fb, Carmona:2001ja, D'Elia:2005ta}. Abelian monopoles are exposed after a gauge fixing and a  ``monopole probe'' is inserted in the lattice: the expectation value of such an observable, usually called $\mu$, should be exactly zero if magnetic symmetry is respected, a value different from zero is instead a clear signature of the presence of an Abelian monopole condensate in the vacuum of the theory. In practice simulations are done by inserting a monopole field in a time slice and measuring a related quantity, the so called $\rho$ parameter defined by
\begin{equation}
  \rho \equiv \frac{\partial}{\partial \beta} \ln \langle \mu \rangle =  \langle S \rangle_S - \langle S_M \rangle_{S_M} 
\end{equation}
in which $S_M$ is the action with the ``monopole probe'' insertion. So $\rho$ is the difference of the action minus the monopole action weighted with the monopole action in the path integral. For details we refer to the cited articles.

The aim of this work is to give a criterion of confinement in $G_2$ group by mean of the disorder parameter $\mu$. Strictly speaking there are no strings in $G_2$, a situation similar to full QCD where string breaking occours. Anyway in full QCD with $N_f = 2$ \cite{D'Elia:2005ta} the operator $\mu$ is nonetheless a good order parameter. So, in the working hypothesis that this is a good observable for confinement, we look for dual superconductivity in the exceptional group $G_2$.

\subsection{G2 group}

We are now going to state some basic facts about the Lie Group $G_2$. Being defined in mathematics as the group of automorphisms of the octonions it can be naturally constructed as a subgroup of the real group $SO(7)$ which has 21 generators and rank 3. To the usual properties of $SO(7)$ matrices
\begin{equation}
  \det \Omega = 1 \qquad \Omega^{-1} = \Omega^{T}
\end{equation}
we have in addition another constraint
\begin{equation}
  T_{abc} = T_{def} \Omega_{da} \Omega_{eb} \Omega_{fc}
  \label{constraint}
\end{equation}
where $T_{abc}$ is a totally antisymmetric tensor whose nonzero elements are (using the octonion basis given by \cite{Cacciatori:2005yb})
\begin{equation}
  T_{123} = T_{176} = T_{145} = T_{257} = T_{246} = T_{347} = T_{365} = 1.
\end{equation}
Equations \ref{constraint} are 7 relations reducing the numbers of generators to 14. This is clearly seen also in this form which is actually used as a check during simulations:
\begin{equation}
  v = \Omega v\qquad  v_k = \Omega_{ji} T_{ijk}\qquad  k = 1 \dots 7
\end{equation}

The fundamental representation of $G_2$ is 7 dimensional and by using the algebra representation of \cite{Cacciatori:2005yb} we can clearly identify an $SU(3)$ subgroup and several $SU(2)$ subgroups, with 6 of them we can cover the whole group, an important property for MC simulations. The first three $SU(2)$ subgroups are the $4 \times 4$ real representation of the group while the remaining three are extremely difficult to simulate with standard techniques. See next section for details on simulations.


The Lie group $G_2$ has rank 2 as $SU(3)$, this implies that its Cartan subgroup, that is the maximal residual abelian subgroup after an abelian gauge fixing, is $U(1) \times U(1)$. Stable monopole solutions are classified according to the homotopy group\footnote{First equality follows from $\pi_1(G_2) = 0$. See for example \cite{Weinberg:1996kr}}:
\begin{equation}
  \pi_2(G_2/U(1)^2) = \pi_1(U(1) \times U(1)) = \mathbb Z \times \mathbb Z 
\end{equation}
i.e. we have two distinct species of monopoles as for $SU(3)$. 

Another interesting homotopy group shows that center vortices are absent in the theory:
\begin{equation}
  \pi_1(G_2/\mathcal C(G_2)) = \pi_1(G_2) = 0
\end{equation}
while for $SU(3)$ for example
\begin{equation}
  \pi_1(SU(3)/\mathbb Z_3) = \mathbb Z_3.
\end{equation}
So $G_2$ is a good playground to study the dual superconductor picture in a theory without center vortices, thus isolating monopole contribution in confinement.

\section{Simulation and Results}
Much work has been devoted to simulations on $G_2$ Yang-Mills theory by the Bern group \cite{Holland:2003jy, Pepe:2005sz}. This group used the so called Fredenhagen-Marcu order parameter \cite{Fredenhagen:1985ft} to demonstrate confinement in $G_2$ theory, at least in the strong coupling limit. This parameter by construction makes sense only at zero temperature. Our proposal concerns a different approach using the magnetic operator $\mu$ of use also in finite temperature simulations.
Here we are going to investigate the thermodynamical and dual superconducting properties of the system. 

First some informations about the MC runs settings. To simulate the gauge theory
\begin{equation}
\mathcal L = \frac{1}{7g^2} {\rm Tr} \, F_{\mu\nu}F_{\mu\nu}
\end{equation}
we used a simple Cabibbo-Marinari update (heat-bath + overrelaxation in a tunable ratio, for every step) for the first three $SU(2)$ subgroups ($4 \times 4$ representation) spanning the $SU(3) \subset G_2$. This simple setting cannot be used for the remaining three subgroups so we make a completely random gauge transformation every $n$ updates (tipically 2) to guarantee the covering of the whole gauge group. Observables measured are the standard plaquette and the Polyakov Loop. To study the thermodynamical properties we simulated several lattices of spatial dimension $N_s = 12, 16, 20, 24, 32$ and $N_t = 6$ ($N_t = 4$ only for the smallest lattice). At the transition we needed histories of more than $10^4$ updates. The code is very fast (using only real algebra), is written using explicitly SSE2 instructions in single precision for the matrix-multiplication core and run on an Opteron farm here in the computer facilities of the Physics Department of the University of Pisa. 

\subsection{Thermodynamics}

To guess where the physical transition should be, we looked at the plaquette and Polyakov loop susceptibilities.

\begin{figure}
\includegraphics[height=.65\textwidth]{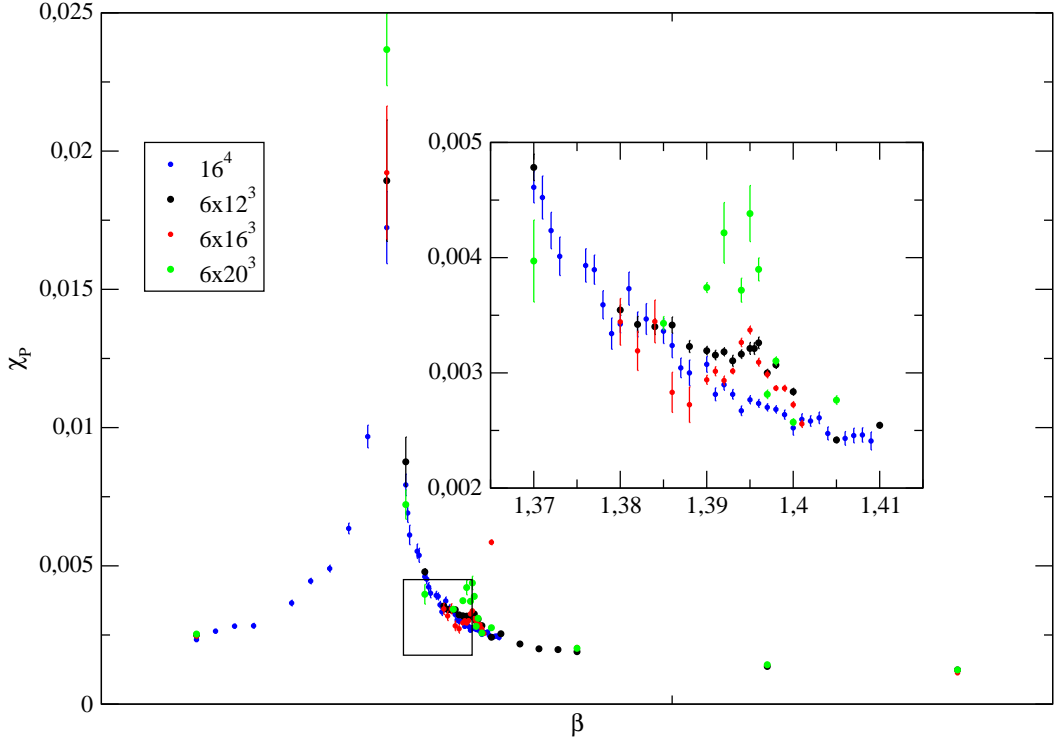}
\caption{Plaquette susceptibility plotted against $\beta$. The peak signals the bulk transition and the points enclosed by the square are the physical transition for the $N_t = 6$ lattices. The transition for $N_t = 4$ is too close to the bulk one and extremely difficult to measure (much of the signal is only ``noise''). Also shown a simulation a T=0 on a $16^4$ lattice (blue points).}
\label{Suscplaq}
\end{figure}

 The first thing to notice is the presence of a bulk transition in the plaquette susceptibility (related to the specific heat), see fig. \ref{Suscplaq}.
 The peak is present in every volume and $N_t$ always at the same $\beta \sim 1.36$. There's no scaling with volume and no movement toward the weak coupling region passing from $N_t = 4$ to $N_t = 6$ as we would expect for a physical transition. This big peak completely overshadows the real physical transition that can be seen as a little peak in the weak coupling region at $\beta \sim 1.395$ for $N_t = 6$ and several spatial volumes. 

\begin{figure}
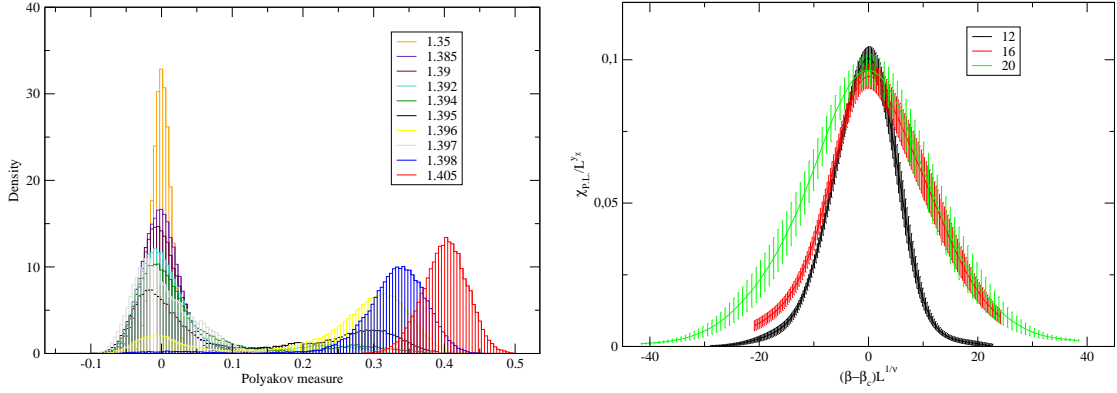

\vspace{5pt}
\includegraphics[height=.34\textwidth]{PolDensity}
\hspace{5pt}
\includegraphics[height=.34\textwidth]{Pol_susc_rescaled}
\caption{Polyakov loop density (on the left) varying $\beta$ in the range from 1.35, the critical coupling of the bulk transition ``$\beta_c$'', to 1.405 where we are in the deconfined phase (all data from the $6\times16^3$ lattice). On the right a rescaling of the Polyakov loop for the smallest lattices assuming a first order transition ($\nu = 1/3$ and $y_\chi = 2.47$ based on a fit on the peak heights). The smallest lattice is evidently not in a scaling region while the other two agree quite well with the hypothesis. Points on the $6\times 16^3$ lattice on the left of $\beta_c$ are probably bad sampled, we just need more statistics.}
\label{Suscpol}
\end{figure}

Polyakov Loop is insensitive to the bulk transition so we used it to detect the position of the physical one, even if it's not an order parameter, see fig. \ref{Suscpol}.
From the behaviour of this observable one can guess a first order transition. An accurate scaling analysis is needed for a conclusive statement.

\subsection{Parameter $\rho$}

\begin{figure}
\vspace{-10pt}
\includegraphics[height=.65\textwidth]{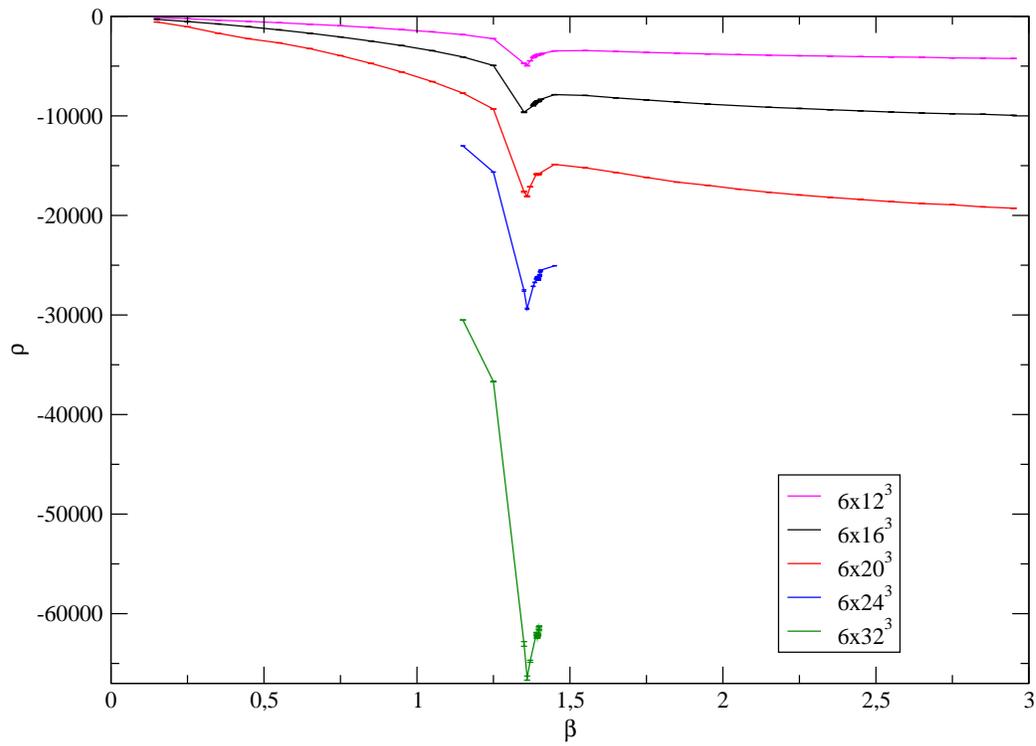}
\caption{Results on the $\rho$ parameter at different lattice volumes, strong dip occours at the bulk transition point.}
\label{Monopole}
\end{figure}

\begin{figure}
\includegraphics[height=.6\textwidth]{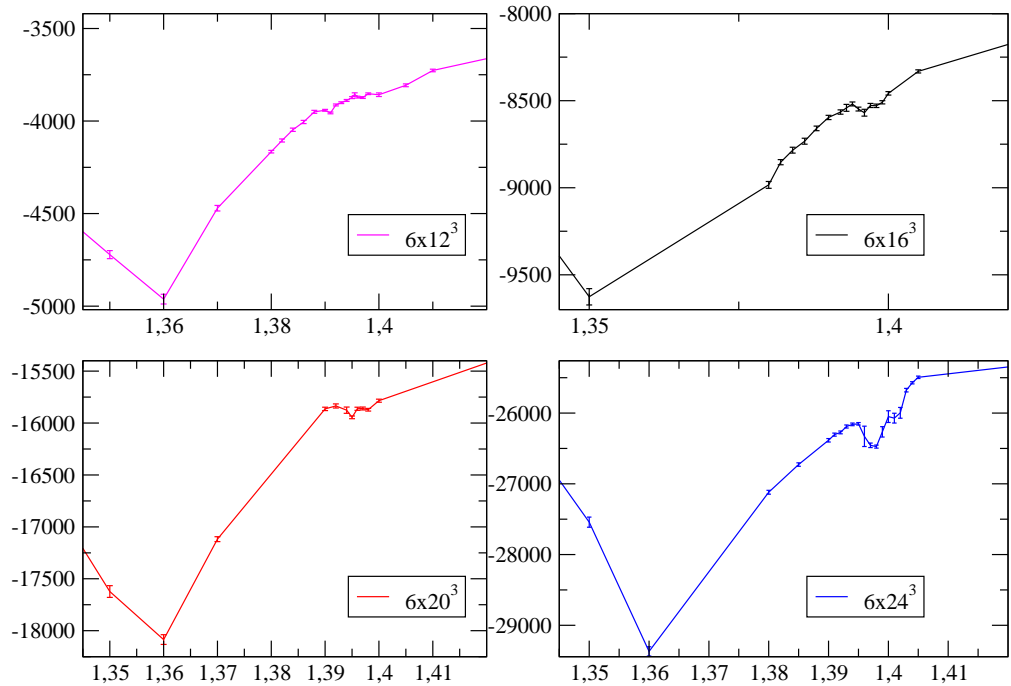}
\caption{Same as before, magnification of the physical transition region.}
\label{Monopole-exp}
\end{figure}

The $\rho$ parameter which, we recall, is defined as the difference between the actions without and with the monopole insertion is expected to be volume independent as $\beta\rightarrow 0$ and to show a dip at the physical transition, see for example \cite{Carmona:2001ja} . Measures of this parameter showed it to be extremely sensitive to the bulk transition with a strong dip at $\beta \sim 1.36$ for all volumes, see fig. \ref{Monopole}.  
Dips at the physical transition are evident only for large volumes $N_s>20$, fig. \ref{Monopole-exp}. The background bulk transition precludes any direct scaling analysis using the monopole operator, moreover volume indepencence before the physical transition can't be seen; key features of the operator are obscured. Another method is needed to study the problem.

A promising method for the $\rho$ susceptibility measure is to renormalize the operator at T=0 subtracting its value $\langle \rho_0 \rangle \equiv \langle \rho(T=0) \rangle$. This is analogous to a renormalization in the free energy density typical in thermodynamical analisies. This way is at an explorative stage.

\subsection{Conclusion}
We are working on the $G_2$ centerless theory in search for confinement by dual superconductivity. Our parameter $\rho$ shows as expected a dip at the transition points but there's an unphysical background that obscures its scaling properties. We are developing a method to deal consistently with this background and find a first order scaling as expected from plaquette and Polyakov Loop Monte Carlo hystories. Also a scaling analysis of the Polyakov Loop susceptibility, even if it is not an order parameter, seems to indicate a first order phase transition. 

We cannot yet conclude that the theory confines but first results, although plagued by the bulk transition background, seems to indicate a positive answer.

\end{document}